\documentclass{main}
\usepackage{textcomp}
\usepackage{fancyhdr}
\usepackage{lastpage}
\usepackage{float,amsmath,amssymb}
\usepackage{graphicx}
\usepackage{subfig}

\def\be{\begin{equation}}
\def\ee{\end{equation}}
\def\bea{\begin{eqnarray}}
\def\eea{\end{eqnarray}}

\newcommand{\Photo}{\includegraphics[height=35mm]{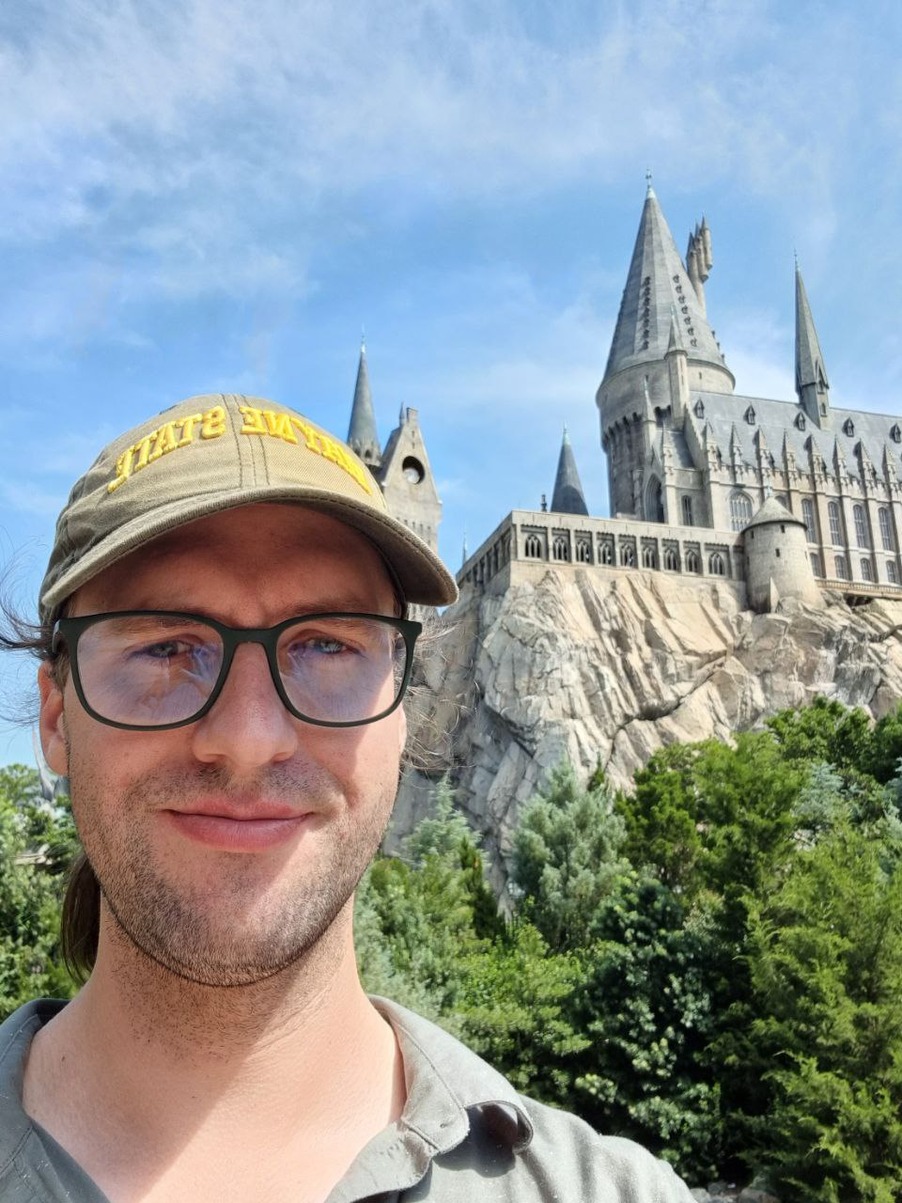}}

\fancypagestyle{firstpagefooter}{%
  \fancyfoot[L]{  \textcopyright \footnotesize This work is an open access article under a Creative Commons Attribution 4.0 International License (https://creativecommons.org/licenses/by/4.0/).}
  \fancyfoot[C]{}  % **Red: This removes the page number from the first page footer**
   
}

\usepackage[dvipsnames]{xcolor}

\begin{document}

\thispagestyle{firstpagefooter}
\title{\Large Nuclear Suppression in Diffractive Vector Meson Production within the Color Glass Condensate Framework}

\author{H.~M\"antysaari$^{1,2}$, \underline{H.~Roch}$^{3,\,}$\footnote{Speaker, email: Hendrik.Roch@wayne.edu}, F.~Salazar$^{4,5,6,7}$, B.~Schenke$^{6}$, C.~Shen$^3$, W. Zhao$^{8,9}$}

\address{
$^1$Department of Physics, University of Jyv\"askyl\"a, P.O. Box 35, 40014 University of Jyv\"askyl\"a, Finland\\
$^2$Helsinki Institute of Physics, P.O. Box 64, 00014 University of Helsinki, Finland\\
$^3$Department of Physics and Astronomy, Wayne State University, Detroit, Michigan 48201, USA\\
$^4$Department of Physics, Temple University, Philadelphia, Pennsylvania 19122, USA\\
$^5$RIKEN-BNL Research Center, Brookhaven National Laboratory, Upton, New York 11973, USA\\
$^6$Physics Department, Brookhaven National Laboratory, Upton, New York 11973, USA\\
$^7$Institute for Nuclear Theory, University of Washington, Seattle, Washington 98195, USA\\
$^8$Nuclear Science Division, Lawrence Berkeley National Laboratory, Berkeley,
California 94720, USA\\
$^9$Physics Department, University of California, Berkeley, California 94720, USA
}

\maketitle\abstracts{
We extend a recent global Bayesian analysis of diffractive $\mathrm{J}/\psi$ production in $\gamma+p$ and $\gamma+\mathrm{Pb}$ collisions within the color glass condensate (CGC) framework to investigate potential modifications of the nucleon structure inside nuclei. 
To this end, we perform fits that allow the effective nucleon structure parameters in Pb nuclei to differ from those of free protons. 
This approach directly addresses the question of whether the proton's spatial gluon distribution at intermediate to large $x$ is modified in the nuclear environment. We compare results obtained with shared and independent nucleon structure parameters and assess the impact on the simultaneous description of $\gamma+p$ data from HERA and the LHC, as well as $\gamma+\mathrm{Pb}$ data from the LHC. 
Our findings show that there is no hint of difference in the nucleon structure beyond those already present in the CGC when embedding nucleons inside a nuclear environment.
}

%\footnotesize DOI: \url{https://doi.org/xx.yyyyy/nnnnnnnn}
%\keywords{one, two, three, four}

\section{Introduction}
\label{sec:intro}
Towards small parton momentum fractions $x$, the gluon density inside hadrons and nuclei rises rapidly when evolved with linear QCD dynamics. 
Eventually, non-linear recombination effects must slow this growth~\cite{Gribov:1983ivg,Mueller:1985wy}, leading to the high-occupancy Color Glass Condensate (CGC) regime~\cite{McLerran:1993ni,McLerran:1993ka,Iancu:2003xm,Morreale:2021pnn,Garcia-Montero:2025hys}. 
Identifying unambiguous signatures of this non-linear QCD regime remains a central challenge in high-energy nuclear physics.

Diffractive vector meson production provides a clean probe of gluons at small~$x$. 
Its cross-section scales with the square of the gluon density at leading order~\cite{Ryskin:1992ui}, and the process is sensitive to the target's spatial structure~\cite{Mantysaari:2020axf}. 
Comparing proton and nuclear targets allows one to study the suppression mechanisms of small-$x$ gluons, which may arise from nuclear shadowing~\cite{Frankfurt:2003qy,Guzey:2013qza,Eskola:2022vpi,Guzey:2024gff} or from gluon saturation dynamics~\cite{Lappi:2013am,SampaiodosSantos:2014puz,Cepila:2017nef,Luszczak:2019vdc,Bendova:2020hbb,Mantysaari:2022sux,Mantysaari:2023xcu}.

CGC-based models constrained by $\gamma+p$ data from HERA successfully describe coherent and incoherent diffractive $\mathrm{J}/\psi$ production~\cite{Mantysaari:2016ykx}, where the incoherent channel revealed the importance of a fluctuating proton substructure~\cite{Mantysaari:2018zdd,Mantysaari:2023qsq}. 
Extending this framework to ultra-peripheral heavy-ion collisions, one nucleus serves as a source of quasi-real photons that probe the other nucleus. 
Recent measurements of coherent $\mathrm{J}/\psi$ production in $\gamma+\mathrm{Pb}$ collisions at the LHC~\cite{ALICE:2023jgu,CMS:2023snh} show stronger nuclear suppression than predicted by CGC calculations constrained solely by HERA data, which typically grow too rapidly with the photon–nucleus center-of-mass energy $W$~\cite{Lappi:2013am,Mantysaari:2022sux,Mantysaari:2023xcu,Mantysaari:2024zxq}. 
This persistent tension has not yet been resolved (see also a recent UPC measurement from ATLAS~\cite{ATLAS:2025aav}, which is not compatible with ALICE data), even after allowing for scale uncertainties or extensions of the non-perturbative proton structure~\cite{Schenke:2024gnj}.

In our recent work~\cite{Mantysaari:2025ltq}, we performed a global Bayesian analysis combining $\gamma+p$ and $\gamma+\mathrm{Pb}$ diffractive $\mathrm{J}/\psi$ data, introducing an overall $K$-factor to absorb model uncertainties from, e.g., the meson wave function and/or missing higher-order effects. 
That study showed that a simultaneous description of both systems is highly constrained, highlighting the difficulty of reconciling the degree of nuclear suppression with parameters consistent with HERA data alone.

In this proceedings contribution, we extend that analysis to directly test whether the effective nucleon structure is modified inside nuclei. 
To this end, we perform Bayesian fits in which the nucleon structure parameters in Pb are allowed to differ from those of free protons. 
This strategy addresses whether the spatial gluon distribution of the proton is altered more than what the CGC framework predicts for a nucleon placed in a nuclear environment. 
We compare fits with shared versus independent nucleon parameters and study their impact on the simultaneous description of $\gamma+p$ and $\gamma+\mathrm{Pb}$ data from HERA and the LHC.

\section{Model}
\label{sec:model}
We compute diffractive $\mathrm{J}/\psi$ production within the Color Glass Condensate (CGC) framework, following Refs.~\cite{Mantysaari:2022sux,Mantysaari:2023xcu,Mantysaari:2025ltq}.
The coherent cross section differential in Mandelstam $t$ is given by
\begin{equation}
\frac{\mathrm{d}\sigma^{\gamma^* + A \to J/\psi + A}}{\mathrm{d}t} = \frac{K}{16\pi} \left| \left\langle \mathcal{A}^{\gamma^*+A \to J/\psi + A} \right\rangle_\Omega \right|^2,
\label{eq:coh}
\end{equation}
where $\langle \cdot \rangle_\Omega$ denotes the average over fluctuating target configurations $\Omega$, and $\mathcal{A}$ is the vector meson production amplitude, obtained from the overlap of the photon and vector meson wave functions convoluted with the dipole–target scattering amplitude~\cite{Kowalski:2006hc,Hatta:2017cte}.
The incoherent cross section, which probes event-by-event fluctuations of the target structure, is determined from the variance of the amplitude:
\begin{equation}
\frac{\mathrm{d}\sigma^{\gamma^* + A \to J/\psi + A^\ast}}{\mathrm{d}t} = \frac{K}{16\pi} \left( \left\langle \left| \mathcal{A}^{\gamma^\ast+A \to J/\psi + A^*} \right|^2 \right\rangle_\Omega - \left| \left\langle \mathcal{A}^{\gamma^\ast+A \to J/\psi + A^*} \right\rangle_\Omega \right|^2 \right).
\label{eq:inc}
\end{equation}

The dipole amplitude is calculated from Wilson lines sampled using the McLerran–Venugopalan (MV) model~\cite{McLerran:1993ni} at $x_\mathbb{P}=0.01$, and subsequently evolved to smaller $x_\mathbb{P}$ with the  Jalilian-Marian, Iancu, McLerran, Weigert, Leonidov and Kovner (JIMWLK) equation on an event-by-event basis~\cite{Mueller:2001uk,Lappi:2012vw}.
The proton substructure is described in terms of fluctuating hot spots~\cite{Mantysaari:2018zdd}, with shape, density, and fluctuation parameters listed in the first block of Table~\ref{tab:parameters}.
Parameters governing the JIMWLK evolution are shown in the second block of the table.  

It has been shown that introducing a phenomenological $K$ factor in Eqs.~\eqref{eq:coh}–\eqref{eq:inc} --- to account for uncertainties from the vector meson wave function and higher-order corrections --- improves the simultaneous description of $\gamma+p$ and $\gamma+\mathrm{Pb}$ data~\cite{Mantysaari:2025ltq}.
Since our goal here is to investigate potential modifications of the nucleon structure in the nuclear environment, we fix the $K$ factor to unity.  

In this work, we extend our previous analysis~\cite{Mantysaari:2025ltq} by allowing the effective nucleon structure parameters in Pb nuclei to differ from those of free protons.
Concretely, we introduce system-dependent nucleon sizes $B_G^{\rm Pb}$ and $B_{G}^{p}$ as well as hot spot sizes $B_{G,q}^{\rm Pb}$ and $B_{G,q}^{p}$.
Predictions are obtained by sampling 25 parameter sets from the resulting posterior distribution, thereby propagating parameter uncertainties into the calculated cross sections.

\begin{table*}[htb!]
    \caption{Summary of model parameters and their prior ranges. Block~1 corresponds to the baseline setup used in previous studies~\protect\cite{Mantysaari:2022sux,Mantysaari:2023xcu,Mantysaari:2025ltq}, and Block~2 to JIMWLK evolution parameters.}
    \label{tab:parameters}
    \begin{tabular}{l|l|l|l}
    \hline\hline
    Block & Parameter & Description & Prior range \\
    \hline
    1 & $m\;[\mathrm{GeV}]$ & Infrared regulator &  $[0.02,1.2]$ \\
    & $B_G\;[\mathrm{GeV}^{-2}]$ & Proton size & $[1,10]$ \\
    & $B_{q}\;[\mathrm{GeV}^{-2}]$ & Hot spot size & $[0.05,3]$ \\
    & $\sigma$ & Magnitude of $Q_s$ fluctuations & $[0,1.5]$ \\
    & $Q_s/(g^2\mu)$ & Ratio of color charge density to saturation scale & $[0.05,1.5]$ \\
    \hline
    2 & $m_{\mathrm{JIMWLK}}\;[\mathrm{GeV}]$ & Infrared regulator & $[0.02,1.2]$ \\
    & $\Lambda_{\mathrm{QCD}}\;[\mathrm{GeV}]$ & Spatial $\Lambda_{\rm QCD}$ & $[0.0001,0.28]$ \\
    \hline\hline
    \end{tabular}
\end{table*}

\section{Results}
\label{sec:results}
\begin{figure*}[t!]
    \includegraphics[width=\textwidth]{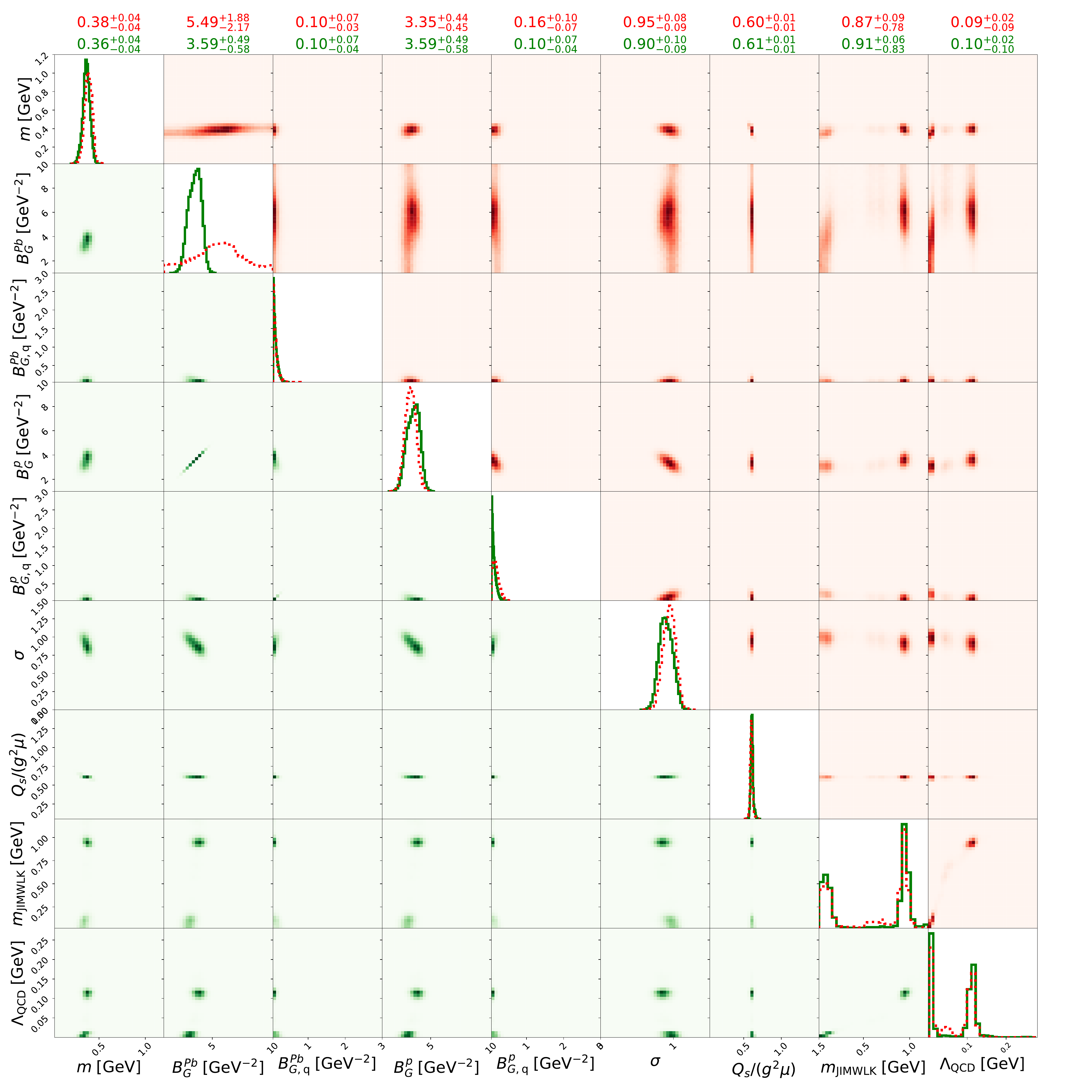}
    \caption{Posterior distributions obtained in the global analysis with (dotted, upper corner) and without (solid, lower corner) the model extension introducing separate nuclear structure parameters $B_G$ and $B_{G,q}$ for the two systems. The numbers at the top denote the median values along with their corresponding 90\% credible intervals.}
    \label{fig:posteriorNuclStruct}
\end{figure*}

In Fig.~\ref{fig:posteriorNuclStruct} we show the posterior distribution from Ref.~\cite{Mantysaari:2025ltq}, referred to in the following as the \textit{standard} setup (solid lines, lower corner).
The \textit{extended} setup, which introduces separate nuclear structure parameters for the $\gamma+p$ and $\gamma+\mathrm{Pb}$ systems, is shown with dotted lines in the upper corner.  

Most marginalized posterior distributions remain largely unchanged in the extended setup.
We observe a slight increase in the magnitude of the $Q_s$ fluctuations, $\sigma$, compensated by a decrease in $B_G^p$, reflecting the anti-correlation between the two parameters observed previously~\cite{Mantysaari:2025ltq}.
The most pronounced difference is the broadening of $B_G^{\rm Pb}$, which arises because the $\gamma+\mathrm{Pb}$ system constrains only the overall nuclear size, but not the individual nucleon size very well.  

For the JIMWLK parameters, only marginal changes are observed.
This indicates that the evolution speed remains essentially unmodified in the extended setup, already suggesting that the extension does not improve the simultaneous description of the two systems.  

\begin{figure*}[tb]
  \centering
  \subfloat[$\gamma+p\to J/\psi+p$ cross section as a function of center-of-mass energy compared to ALICE~\protect\cite{ALICE:2014eof,ALICE:2018oyo}, H1~\protect\cite{H1:2005dtp,H1:2013okq}, ZEUS~\protect\cite{ZEUS:2002wfj}, and LHCb~\protect\cite{LHCb:2018rcm,LHCb:2024pcz} data. Datapoints with open markers are not included in the fit.]{%
    \includegraphics[width=0.48\textwidth]{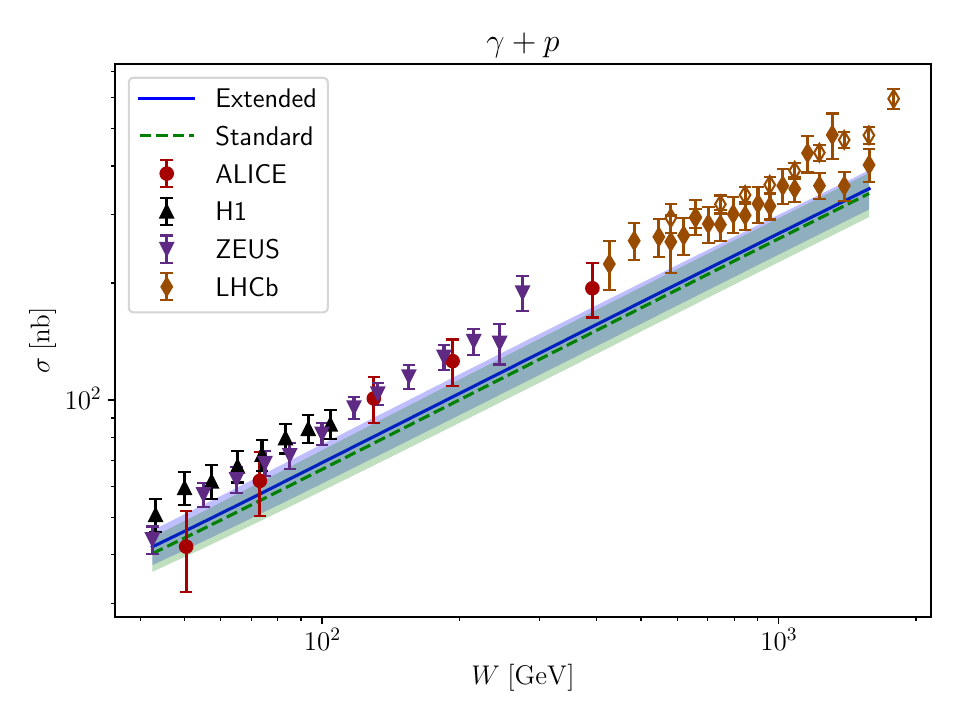}%
    \label{fig:gammap}%
  }
  \hfill
  \subfloat[$\gamma+\mathrm{Pb}\to J/\psi+\mathrm{Pb}$ cross section as a function of center-of-mass energy compared to ALICE~\protect\cite{ALICE:2023jgu} and CMS~\protect\cite{CMS:2023snh} data.]{%
    \includegraphics[width=0.48\textwidth]{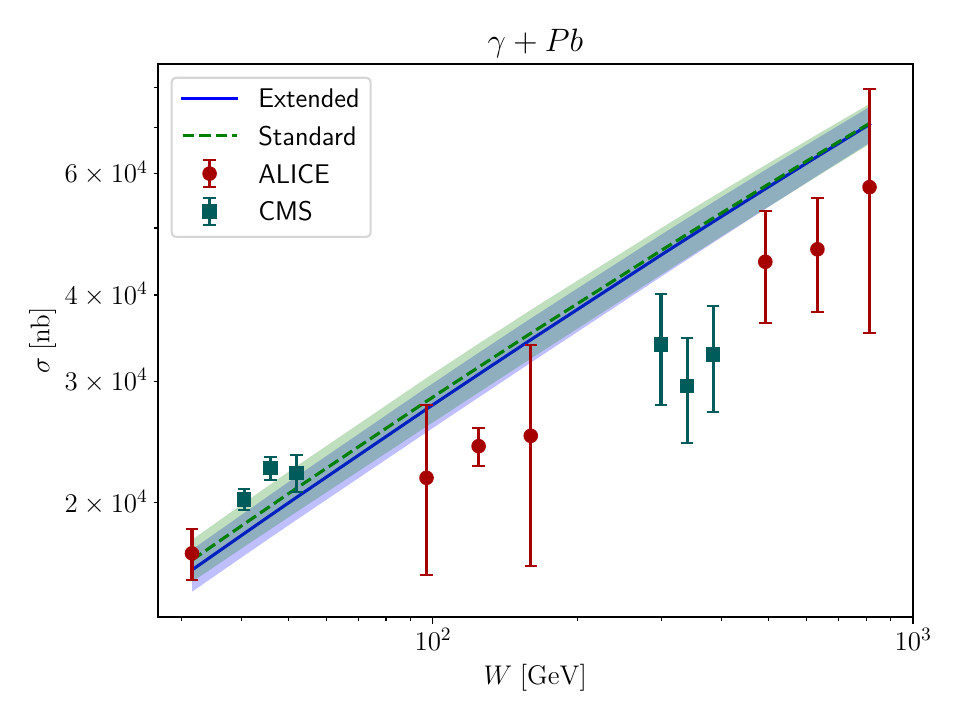}%
    \label{fig:gammaPb}%
  }
  \hfill
  \subfloat[Coherent (smaller $|t|$) and incoherent (larger $|t|$) $J/\psi$ spectra in $\gamma+p$ compared to H1 data~\protect\cite{H1:2013okq}.]{%
    \includegraphics[width=0.48\textwidth]{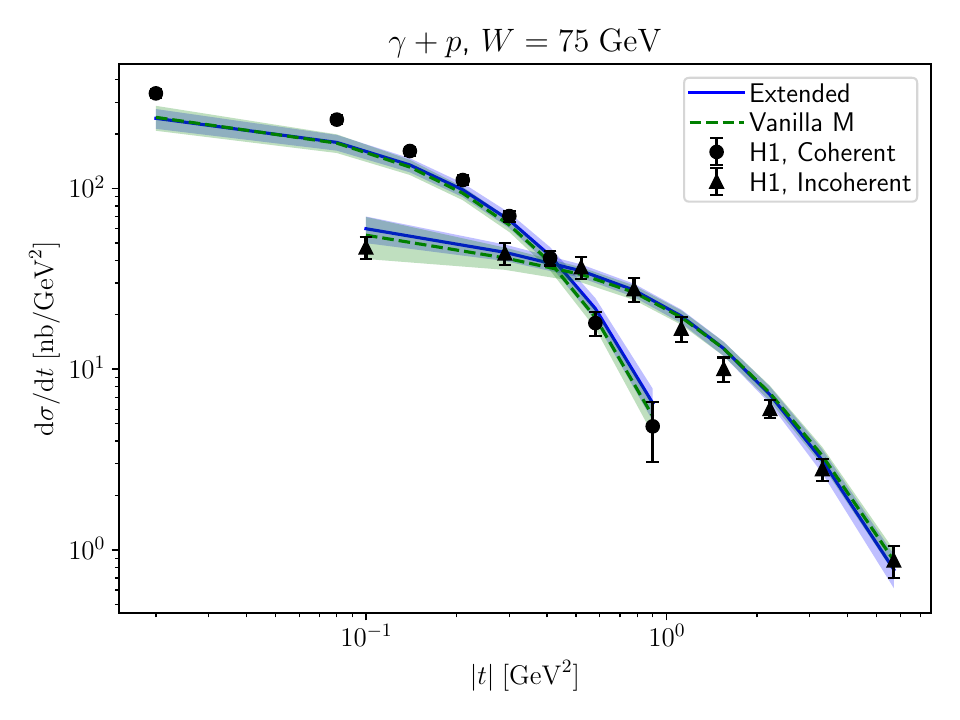}%
    \label{fig:gammap_t}% 
    
  }
  \hfill
  \subfloat[Coherent (smaller $|t|$) and incoherent (larger $|t|$) $J/\psi$ spectra in $\gamma+\mathrm{Pb}$ compared to ALICE data~\protect\cite{ALICE:2021tyx,ALICE:2023gcs}.]{%
    \includegraphics[width=0.48\textwidth]{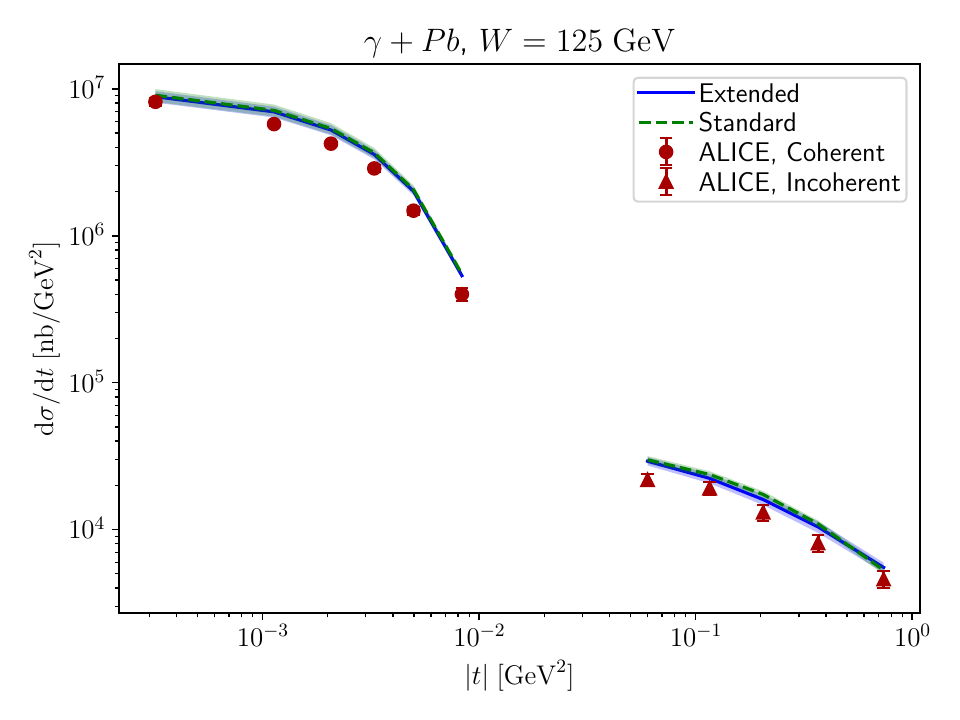}%
    \label{fig:gammaPb_t}%
  }
  \caption{Integrated $W$ dependent and $|t|$-differential cross sections from the fit containing $\gamma+p$ and $\gamma+\mathrm{Pb}$ data in the standard parameter setup (dashed) and including separate nuclear structure parameters $B_G$ and $B_{G,q}$ (full) the uncertainty bands indicate the 68\% credible intervals from 25 posterior sample runs of the model.}
  \label{fig:obs_comparison}
\end{figure*}

In Figs.~\ref{fig:gammap}–\ref{fig:gammaPb_t} we compare model predictions for the $W$-dependent integrated cross sections and the $|t|$-differential cross sections between the standard setup (dashed) and the extended setup (solid).  
Across all observables, we find no significant differences between the two parameterizations, and both remain consistent within uncertainties.  
As anticipated, the extended setup does not provide an improved description of the experimental data.

To understand in more detail the effect of a change in the parameters on the final observables, we investigate the local region of the parameter space constrained by the Bayesian inference for the extended model.
For this, we compute the response coefficients between experimental observables and the model's parameters.
We follow the approach introduced in Refs.~\cite{Sangaline:2015isa,Jahan:2024wpj} and define an ensemble-averaged response coefficient matrix,
\begin{align}
    \mathcal{R}_{ai} \equiv \left\langle \frac{\partial y_a}{\partial \theta_i} \right\rangle_\mathrm{post} = \sum_j \langle \delta y_a \delta \theta_j \rangle_\mathrm{post} \langle \delta \theta_j \delta \theta_i \rangle_\mathrm{post}^{-1},
    \label{eq:responseMatrix}
\end{align}
where the index $a$ represents the individual experimental observables and $i$ the parameters of the model.
The posterior average $\langle \cdot \rangle_\mathrm{post}$ is employed to evaluate the model response in the neighborhood of the experimental data.
To define the response coefficients dimensionless, we rescale observables and parameters as $\delta y_a = (y_a(\theta_j) - \bar{y}_a)/\sigma_{y_a}$ and
$\delta \theta_i = (\theta_i - \bar{\theta}_i)/\sigma_{\theta_i}$, where $\bar{y}_a$ ($\bar{\theta}_i$) and $\sigma_{y_a}$ ($\sigma_{\theta_i}$) denote the posterior mean and standard deviation of $y_a$ ($\theta_i$).
The expression on the right-hand side of Eq.~\eqref{eq:responseMatrix} should be understood as an ordinary matrix multiplication.

\begin{figure*}[t!]
    \centering
    \includegraphics[width=0.32\textwidth]{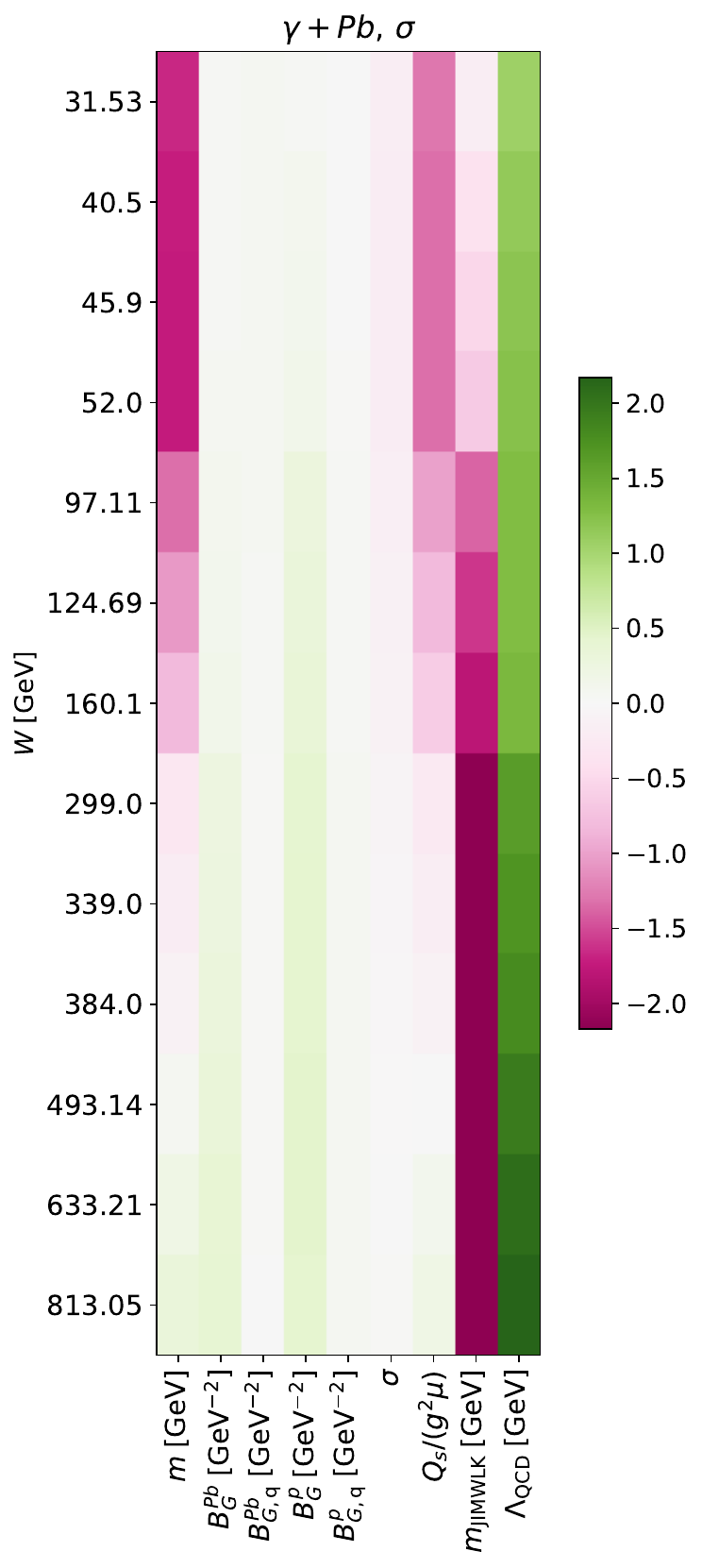}
    \includegraphics[width=0.32\textwidth]{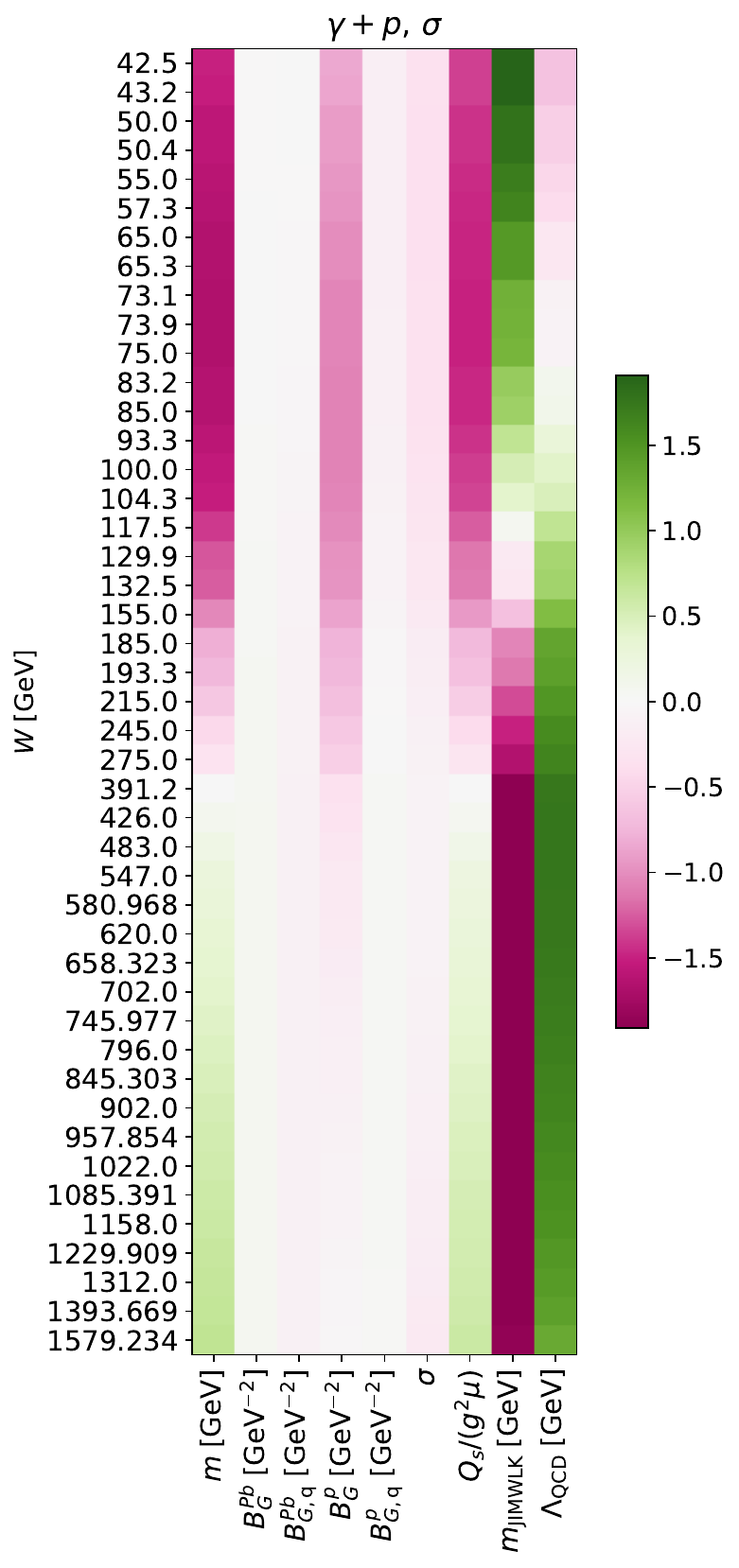}
    \caption{Response matrices $\mathcal{R}_{ai}$ for the integrated $\gamma+\mathrm{Pb}$ (left) and $\gamma+p$ (right) cross sections, averaged over the posterior.}
    \label{fig:sensitivity_int}
\end{figure*}

Figure~\ref{fig:sensitivity_int} shows the response coefficient matrices for the integrated cross sections of the two systems.  
We find a strong sensitivity to the infrared regulator $m$ at small $W$ in both $\gamma+p$ and $\gamma+\mathrm{Pb}$, as it determines the initial condition for the JIMWLK energy evolution.  
In addition, the cross sections at high-$W$ are highly sensitive to the JIMWLK parameters $m_{\rm JIMWLK}$ and $\Lambda_{\rm QCD}$, which govern the evolution speed.  
These two parameters are anti-correlated, acting in the same direction at large $W$ for both systems.  
Interestingly, in the $\gamma+p$ case this anti-correlation changes sign at small $W$.  
We also observe a noticeable sensitivity to the proton size $B_G^p$ at low $W$ for $\gamma+p$ collisions.  

\begin{figure*}[t!]
    \centering
    \includegraphics[width=0.24\textwidth]{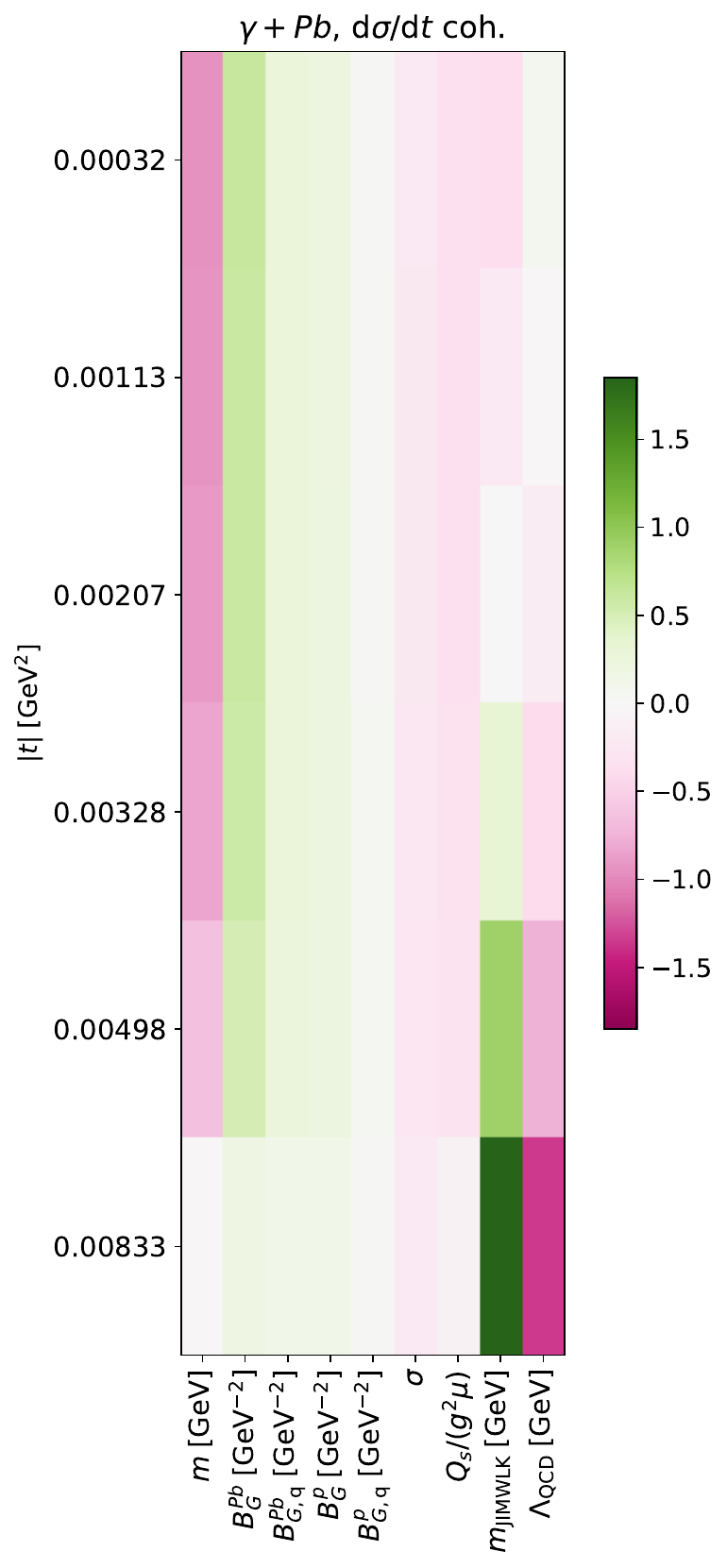}
    \includegraphics[width=0.24\textwidth]{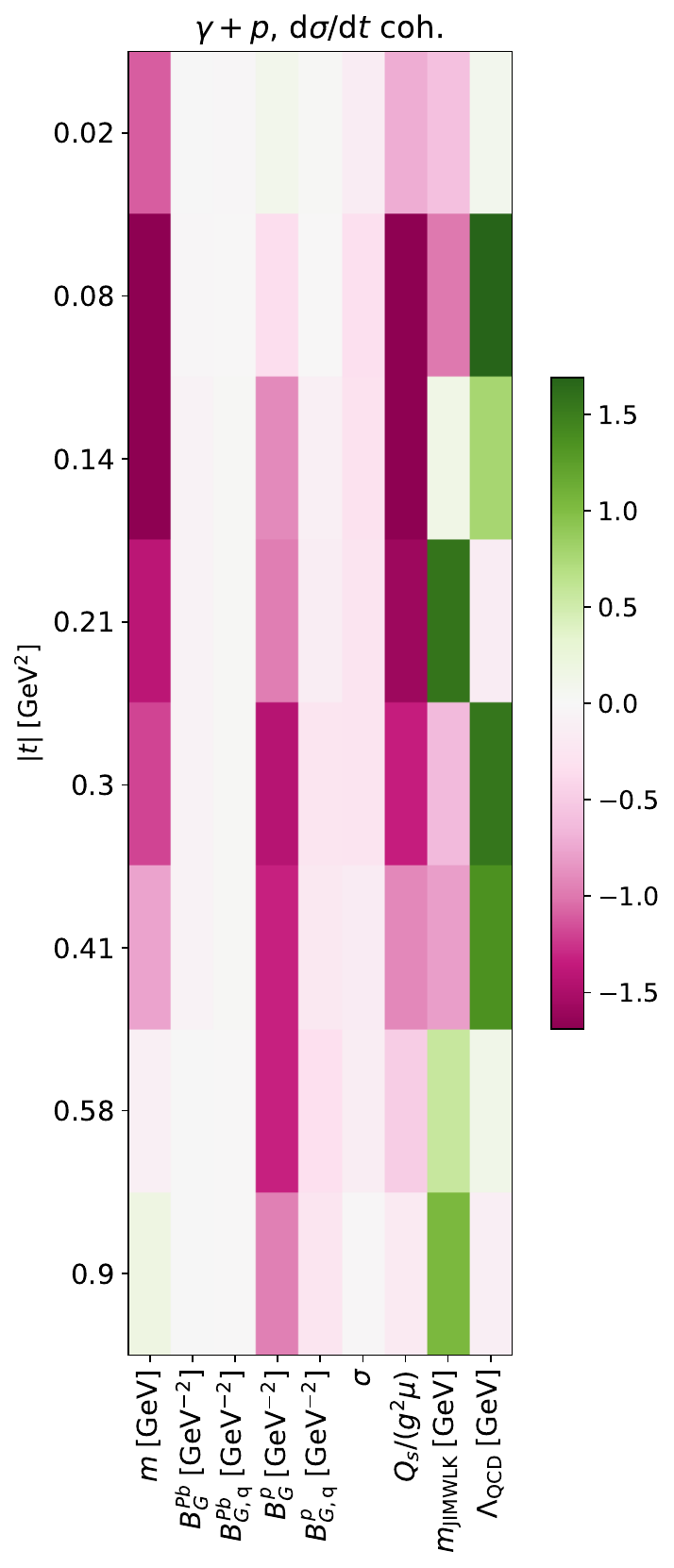}
    \includegraphics[width=0.24\textwidth]{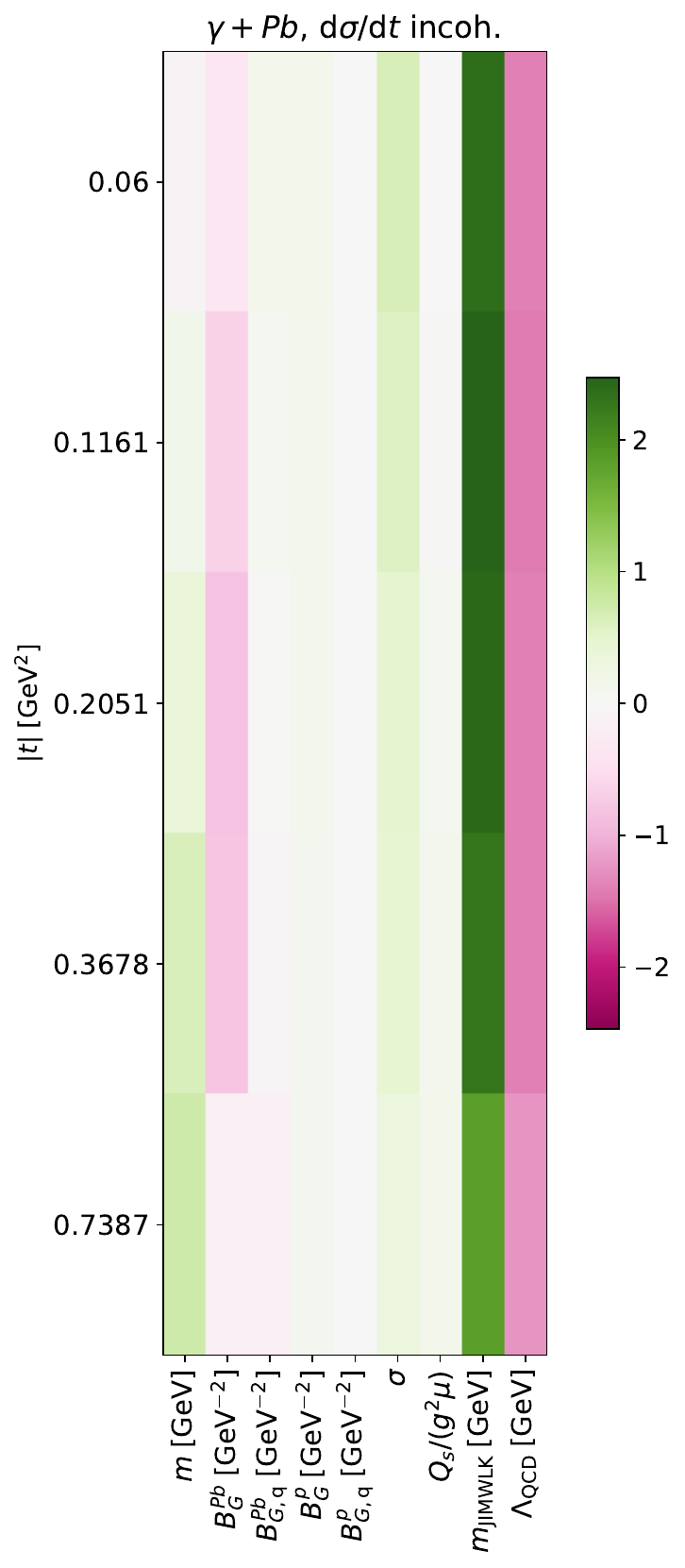}
    \includegraphics[width=0.24\textwidth]{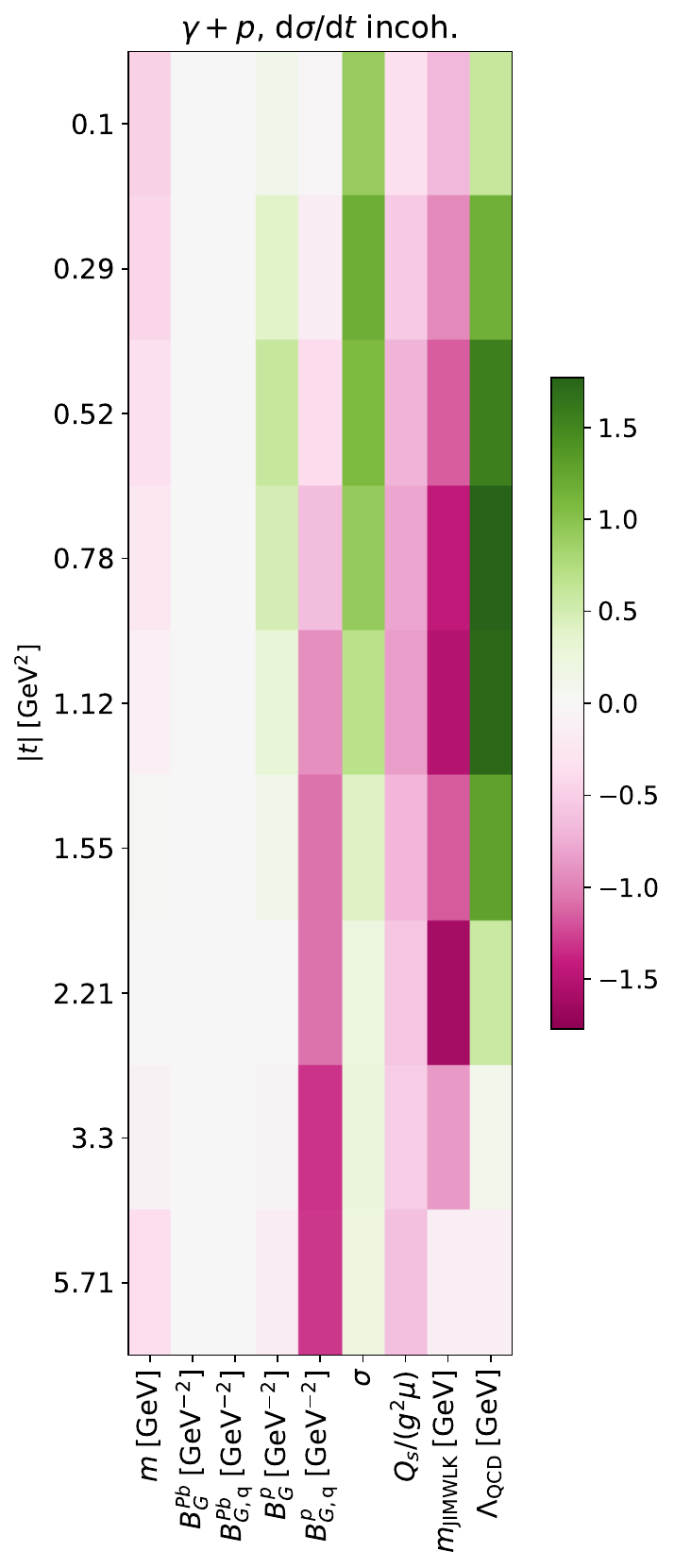}
    \caption{Response matrices $\mathcal{R}_{ai}$ for the $|t|$-differential cross sections: coherent $\gamma+\mathrm{Pb}$ (far left) and $\gamma+p$ (center left), as well as incoherent $\gamma+\mathrm{Pb}$ (center right) and $\gamma+p$ (far right). Results are averaged over the posterior.}
    \label{fig:sensitivity_diff}
\end{figure*}

Figure~\ref{fig:sensitivity_diff} presents the response coefficient matrices for the $|t|$-differential cross sections.  
The left two panels correspond to the coherent case: in $\gamma+\mathrm{Pb}$ (far left), we see only a weak response of the low-$|t|$ cross section to the nucleon size $B_G^{\rm Pb}$, while in $\gamma+p$ (center left) the sensitivity at large $|t|$ is much stronger for $B_G^p$.
This aligns with the constraints we have seen in Fig.~\ref{fig:posteriorNuclStruct}, where the distribution for $B_{G}^{\rm Pb}$ is less constrained than the one for $B_G^p$.
We also observe sensitivity to the infrared regulator $m$ at small $|t|$, which affects the size and shape of the target.
The right two panels show the incoherent case: $\gamma+p$ (far right) exhibits a strong response at large $|t|$ to the hot spot size $B_{G,q}^p$, whereas the $\gamma+\mathrm{Pb}$ spectrum (center right) does not extend to sufficiently large $|t|$ to probe this sensitivity.  
Finally, the decreasing sensitivity to $\sigma$ with increasing $|t|$ in the incoherent $\gamma+p$ cross section has already been observed in Ref.~\cite{Mantysaari:2016jaz}.

\section{Conclusion}
\label{sec:conclusion}
We have extended a previous Bayesian inference study~\cite{Mantysaari:2025ltq} to address whether the nucleon structure is modified beyond what the CGC predicts for a nucleon embedded in a nuclear environment.
Our analysis finds no evidence for such additional modification. 
Furthermore, the extended model does not lead to an improved simultaneous description of the $\gamma+\mathrm{Pb}$ and $\gamma+p$ systems. 
Instead, the inclusion of a simple $K$-factor within the standard setup continues to provide the best overall agreement with experimental data. 

In addition, we have examined the sensitivity of the observables in the posterior region of parameter space. 
As expected, the integrated cross sections show strong dependence on the JIMWLK energy evolution parameters, while the $|t|$-differential $\gamma+p$ observables are most sensitive to nucleon structure parameters. 
For the incoherent $\gamma+\mathrm{Pb}$ cross section, the sensitivity to the hot spot size is suppressed because the experimental data do not extend to the same large-$|t|$ region as in the $\gamma+p$ case, where this sensitivity becomes apparent.

\section*{Acknowledgments}
H.M. is supported by the Research Council of Finland, the Centre of Excellence in Quark Matter, and projects 338263 and 359902, and under the European Research Council (ERC, grant agreements No. ERC-2023-101123801 GlueSatLight and No. ERC-2018-ADG-835105 YoctoLHC).
This work is supported by the U.S. Department of Energy, Office of Science, Office of Nuclear Physics, under DOE Contract No.~DE-SC0012704 (B.P.S.), DOE Award No. DE-SC0021969 (C.S.) and DE-SC0024232 (C.S. \& H.R.), and within the framework of the Saturated Glue (SURGE) Topical Theory Collaboration (F.S., B.P.S., W.Z.).
H.R. and W.Z. were supported in part by the National Science Foundation (NSF) within the framework of the JETSCAPE collaboration (OAC-2004571).
C.S. acknowledges a DOE Office of Science Early Career Award.
This research was done using resources provided by the Open Science Grid (OSG)~\cite{Pordes:2007zzb,Sfiligoi:2009cct}, which is supported by the National Science Foundation awards \#2030508 and \#2323298. 
F.S. is supported by the Laboratory Directed Research and Development of Brookhaven National Laboratory and RIKEN-BNL Research Center. 
Part of this work was conducted while F.S. was supported by the Institute for Nuclear Theory of the U.S. DOE under Grant No. DE-FG02-00ER41132. 
Part of the numerical simulations presented in this work were performed at the Wayne State Grid, and we gratefully acknowledge their support.
The content of this article does not reflect the official opinion of the European Union, and responsibility for the information and views expressed therein lies entirely with the authors.

\section*{References}
\bibliography{bib}

\end{document}